\documentclass[APS,prl,twocolumn,amsmath,amssymb,nofootinbib,preprintnumbers,floatfix,reprint]{revtex4-2}
\usepackage{graphicx}% Include figure files
\usepackage{dcolumn}% Align table columns on decimal point
\usepackage{xcolor}
\usepackage{bm}% bold math
\newcommand{\bq}{\boldsymbol q}

\newcommand{\bx}{\boldsymbol x}

\newcommand{\bk}{\textbf{k}}

\usepackage[normalem]{ulem}

\newcommand{\cmfast}{\texttt{21cmFAST} }

\newcommand{\opt}{\texttt{OPT} }

\begin{document}

\preprint{APS/123-QED}

\title{Local primordial non-Gaussianity from `zero-bias' 21cm radiation during reionization}
\author{Nickolas Kokron}
\email{kokron@ias.edu}
\affiliation{School of Natural Sciences, Institute for Advanced Study, 1 Einstein Drive, Princeton, NJ, 08540, USA}
\affiliation{Department of Astrophysical Sciences, Princeton University, 4 Ivy Lane, Princeton, NJ, 08544, USA}
\date{\today}

\begin{abstract}
We revisit the potential of 21cm radiation fluctuations, during the epoch of reionization, in constraining the amplitude of local primordial non-Gaussianity (PNG) $f_{\rm NL}^{\rm loc}$. There generically exists an epoch at which the linear bias of the 21cm field crosses zero, independent of the precise astrophysics of reionization. This epoch implies the 21cm radiation is a natural \emph{zero-bias tracer} in the sense of Castorina et al.~\cite{Castorina_2018}. We identify new noise-like contributions which directly compete with the zero-bias effect, but which should be mitigated through sophisticated analysis techniques such as field-level reconstruction. These noise-like terms act to hinder the constraining power on local PNG of the brightness temperature fluctuations, making $\sigma (f_{\rm NL}^{\rm loc}) \leq 1$ unachievable even in simplified forecasts. We show that analyses which can reach the `sampling noise' floor for this tracer and harness its full power can potentially unlock a 10-fold reduction in error bars, even in the presence of large-scale cuts from foregrounds. The potential of this epoch motivates searching for it in future 21cm surveys, along with developing analysis techniques that can reach the noise floor required for the zero-bias epoch to saturate Fisher information.

\end{abstract}
\maketitle
\section*{Introduction} 
The 21cm hyperfine transition of neutral Hydrogen is an important tracer of the distribution of gas in the Universe across cosmic time. Measuring the global evolution of the brightness temperature of 21cm radiation and its anisotropies has the potential to revolutionize our understanding of the epoch of cosmic reionization, as well as the `dark ages' at even higher redshifts. Many surveys are actively looking for such signals and many others will start collecting data in in the future~\cite{DeBoer_2017, Wayth_2018, Weltman_2020}. High redshift maps of 21cm fluctuations will probe immense cosmic volumes, and have the potential to be highly sensitive probes of both standard cosmology and extensions to it. \par  
In the limit of small optical depth, at a given 3D position in the Universe $\bx$ and redshift $z$, the differential brightness temperature measured by a 21cm survey can be approximated as depending on \cite{LoebFurlanetto}
\begin{align}
\label{eqn:dtb}
    \delta T_b (\bx,z) = &T_0 (z) (1 + \delta_m(\bx)) x_H(\bx) \left [ 1 - \frac{T_{\rm CMB}}{T_s (\bx)} \right] \\
    \nonumber &\times \left ( \frac{H(z)}{(1+z) \partial_{r_\parallel} v_\parallel} \right ),
\end{align}
where $\delta_m (\bx)$ is the dark matter density contrast, $x_H(\bx)$ is the fraction of neutral hydrogen, $T_s$ is the spin temperature of neutral hydrogen, $dv_\parallel / dr_\parallel$ is the local velocity gradient of the neutral hydrogen gas cloud (which causes redshift-space distortions in the signal) and $T_0$ is a global temperature given by~\cite{cruz2024billionyearssecondseffective} 
\begin{equation}
    T_0 (z) = 34 {\rm mK} \left ( \frac{1+z}{16} \right)^{1/2} \left ( \frac{\Omega_b h^2}{0.022} \right ) \left ( \frac{\Omega_m h^2}{0.14} \right )^{-1/2}.
\end{equation}
The fluctuations in the density of dark matter are presumed to be seeded by an inflationary epoch in the Universe. The exact nature of this inflation is not well known and finding signatures of this epoch is one of the main goals of modern Cosmology. Models of inflation which go beyond the fiducial `single-field slow roll' paradigm often seed local-type primordial non-Gaussianity (PNG) in the distribution of curvature perturbations. For a Gaussian potential $\phi$ the matter fluctuations will evolve according to a Poisson equation with a potential $\Phi$ 
\begin{equation}
    \Phi (\bx) = \phi(\bx) + f_{\rm NL}^{\rm loc} \left ( \phi^2 (\bx) - \langle \phi^2 \rangle \right ),
\end{equation}
and $f_{\rm NL}^{\rm loc}$ is the parameter that controls the amplitude of PNG and is dependent on the microphysics of inflation. Future surveys are targeting $\sigma(f_{\rm NL}^{\rm loc}) \sim 1$ as an important threshold that distinguishes single-field from multi-field inflation models~\cite{ferraro2022snowmass2021cosmicfrontierwhite}. \par 
In the presence of PNG, biased tracers of the dark matter distribution (such as dark matter haloes and the galaxies inside them) have their number densities modulated by the presence of this primordial potential~\cite{McDonald_2008,Assassi_2015}
\begin{equation}
\label{eqn:tracerX}
    n_X (\bx) \approx \bar{n}_X \left ( 1 + b_1 \delta_m (\bx) + f_{\rm NL} b_\phi \phi(\bx) + \cdots + \epsilon(\bx) \right ),
\end{equation}
with $\bar{n}_X$ being the average number density of a tracer $X$ and $\epsilon(\bx)$ characterizing scatter in this statistical relation. The modulation of galaxy number density by the presence of large-scale potential fluctuations induces a \emph{scale-dependent bias} in the power spectrum of density contrasts $(\delta_X (\bx) \equiv n_X (\bx) / \bar{n}_X - 1)$~\cite{Dalal_2008}
\begin{equation}
    P_{XX} (k) \underset{k \to 0}{\propto} 2 f_{\rm NL} b_1 b_\phi \alpha(k) P(k) + (f_{\rm NL} b_{\phi} \alpha(k) )^2 P(k),
\end{equation}
where $\alpha(k)$ is the Fourier-space transfer function for the Poisson equation $\phi (\bk) = \alpha(k)\delta (\bk)$ and scales as $\alpha(k) \propto k^{-2}$. It was shown in Ref.~\cite{Castorina_2018} that, if one could measure the large-scale power spectrum of a tracer which possesses $b_1 \approx 0$, under certain conditions, constraints on PNG would be significantly enhanced compared to the traditional approach of selecting samples with high biases and large volumes (such as quasars~\cite{Slosar_2008, mueller2021clusteringgalaxiescompletedsdssiv, cagliari2023optimalconstraintsprimordialnongaussianity}). Ref~\cite{Castorina_2018} was interested in constructing a zero-bias tracer out of galaxy catalogs. They rely on carefully constructed catalogs which exploit the environmental dependence of the linear galaxy bias, $b_1$, in order to achieve an effective zero-bias. \par 
The sources of reionization are contained in collapsed objects, and thus the 21cm brightness temperature will also exhibit this scale-dependent signal. Past studies~\cite{Joudaki_2011,D_Aloisio_2013, Lidz_2013} of PNG and reionization focused on $\langle z \rangle \sim 8$ and neglected the redshift evolution of this signal. \par
21cm radiation from reionization is a \emph{natural} zero-bias tracer: 21cm radiation traces the density of neutral hydrogen in the Universe which, at high redshifts prior to reionization, is approximately co-located with the dark matter. This implies, then, that there exists a positive correlation between $\delta T_{b}$ and the matter distribution. Late into reionization, most of the Universe will have no 21cm signal. Only regions far from the sites of reionization will be bright in 21cm. The signal will be anti-correlated with large-scale structure, and thus its bias must be negative. At some point, the correlation between the brightness temperature and the dark matter density field must cross zero. \par 
The purpose of this work is to uncover the conditions under which 21cm radiation at this `zero-bias' epoch becomes a powerful probe of local PNG. \par
\section*{The zero-bias epoch in semi-analytical models of reionization} We investigate the occurrence of the zero-bias epoch of reionization in the \texttt{21cmFAST} code~\cite{Mesinger_2010,Park_2019}. We specifically use the parameters of the \opt simulation from the Evolution of 21cm-Structure (EOS) project~\cite{Munoz:2021psm}. \opt is a `best-guess' simulation for the evolution of 21cm fluctuations through reionization which tries to reproduce known observational data for UV luminosity functions, the optical-depth to reionization measured by the Planck satellite~\cite{de_Belsunce_2021}, as well as matching the timing of the global 21cm signal to a claimed detection of 21cm absorption by the EDGES collaboration~\cite{Bowman_2018}.
The box length of the simulation is $L_{\rm box}= 1.5 {\rm Gpc}$. Dark matter density perturbations are calculated on a grid with resolution $L_{\rm cell} = 1 $Mpc, and then downgraded to $L_{\rm cell} = 1.5 $Mpc for the astrophysical steps of the code~\footnote{The original \opt simulation computed the dark matter density field at $L_{\rm cell} = 0.5 $Mpc before downsampling to $L_{\rm cell} = 1.5 $Mpc for its astrophysical evolution. Since we are interested in the large-scale signals and not differences induced by very-small-scale clustering, we adopt this slightly worse dark-matter resolution for computational purposes.}. Our simulation is run from $z=35$ to $z=5$ using 38 logarithmically spaced steps (equivalent to the parameter \texttt{z\_step\_factor = 1.05}). \par
\begin{figure}
    \centering
    \includegraphics[width=\columnwidth]{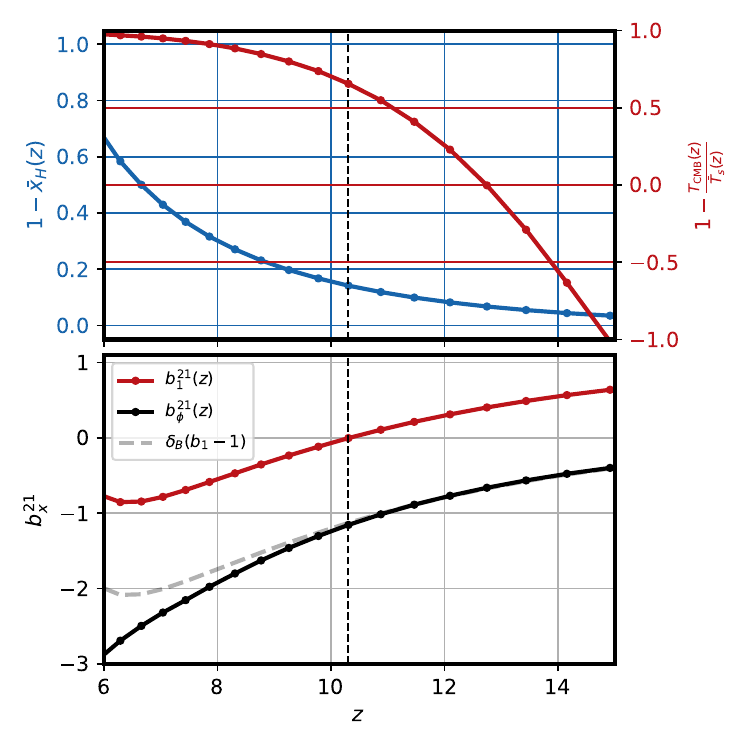}
    \caption{\emph{Top panel}: Evolution of the average ionization fraction in our fiducial model, in blue, as well as evolution of the average spin temperature contribution to the 21cm brightness temperature. The vertical dotted black line indicates the redshift at which the linear bias crosses zero for the 21cm brightness temperature field. It occurs at $z\sim10$, when the Universe is 20\% ionized, and the spin temperature is close to being saturated, justifying neglecting its fluctuations in this work. \emph{Bottom panel:}  Evolution of the linear bias (red) and PNG bias (black) of the 21cm brightness temperature field during reionization. The dashed line shows the commonly-adopted approximation of Eqn.~\ref{eqn:bphi}. }

    \label{fig:b1}
\end{figure}

 
At every snapshot the code outputs the fields $\delta_m(\bx)$, $x_H (\bx)$, $T_s(\bx)$ and $v_i (\bx)$. It is thus possible to reconstruct Eqn.~\ref{eqn:dtb}, or selectively neglect contributions from specific terms. For the purposes of this work we will neglect the contribution from velocity gradients (but their inclusion is simple) and will assume $T_s \gg T_{\rm CMB}$ during the epochs of interest (but note Ref.~\cite{schaeffer2024testingcommonapproximationspredict}). In Fig.~\ref{fig:b1} we show the ionization history and evolution of the average spin temperature for this fiducial simulation. At $z = 10.3$ the Universe is 14\% ionized and $T_s \gtrsim 3 T_{\rm CMB}$.

\section*{Time evolution of the 21cm linear bias}
Under the simplifying assumptions made above, we write the 21cm brightness temperature fluctuation field as 
\begin{equation}
\label{eqn:21cmdtb}
    \delta T_{b}(\bx,z) = T_0 (z) x_H (\bx,z) (1 + \delta_m (\bx,z)).
\end{equation}
We proceed to write fluctuations in the neutral hydrogen fraction field terms of ionization fraction fluctuations, which on large scales admits a symmetries-based biased tracer expansion~\cite{McQuinn_2018, Desjacques_2018}:
\begin{align}
\label{eqn:bubbleexpansion}
    x_H (\bx) \approx 1-\bar{x}_i &\left ( 1 + b_1^x \delta_m (\bx) + b_2^x (\delta_m^2(\bx)- \sigma^2 ) + \cdots \right. \\
    \nonumber &+ \left. \epsilon^x(\bx) \right ).
\end{align}
The operators $\mathcal{O}(\bx) \ni \{\delta_m (\bx), \delta_m^2(\bx), \cdots\}$ encode all possible scalars built from powers of the Hessian of the gravitational potential, and its higher derivatives~\cite{Desjacques_2018}. The field $\epsilon^x(\bx)$ captures the \emph{stochasticity} of this expansion, and encodes the contributions to ionization fraction fluctuations which are uncorrelated with the large-scale operators that characterize this expansion. In certain models, such as excursion-set based models of reionization~\cite{Furlanetto_2004,McQuinn_2005,Lidz_2013}, the bias coefficients $b_n^x$ can be analytically calculated from a mass function for the ionization bubbles. In this work we will extract these coefficients from our simulation. \par 
We define the bias of the 21cm brightness temperature field, $\delta T_b$, through its Fourier cross-correlation with the matter density in the $k\to 0$ limit 
\begin{equation} 
\label{eqn:21cmb1}
b_1^{21} = \lim_{k\to 0} \frac{\langle \delta T_b \delta_m \rangle}{\langle \delta_m \delta_m \rangle}.
\end{equation}
Neglecting velocity gradients and spin-temperature fluctuations, Eqn.~\ref{eqn:21cmdtb} implies the bias expansion of the brightness temperature field is a density-weighting of the bias expansion of the ionization bubbles. 
From Eqns.~\ref{eqn:21cmdtb}, \ref{eqn:bubbleexpansion} and \ref{eqn:21cmb1}, the linear bias of the 21cm field is given by 
\begin{equation}
\label{eqn:linbiaspred}
    b_1^{21}(z) = \bar{x}_H(z) - (1 - \bar{x}_H(z))\left [b_1^x (z)  - 2 \sigma^2 (z) b_2^x(z) \right ].
\end{equation}
 The first two terms recover the result for the linear bias of the 21cm field given in \cite{McQuinn_2018}. The second term arises from the 21cm field being a product of two fields, but at high redshifts and for a bias expansion defined on suitably large scales $\sigma^2 (z)$ is small and can be neglected.\par 
Eqn.~\ref{eqn:linbiaspred} makes it clear that there can be an epoch where the 21cm brightness temperature becomes a zero-bias tracer. The existence of this epoch was previously noted in Ref.~\cite{McQuinn_2018}, where for three different reionization simulations the zero-crossing occurs at $\bar{x}_H > 0.8$. In the \texttt{Thesan-2} reionization simulations~\cite{Qin_2022} the linear bias of the 21cm field crosses zero between $\bar{x}_H \sim 0.8-0.9$ for all of their reionization scenarios. Similarly, from Fig.~\ref{fig:b1} we see that zero-crossing for the scenario considered in this work occurs at $\bar{x}_H = 0.86$, despite completely different physics for reionization than those used in Ref.~\cite{McQuinn_2018} and \texttt{Thesan-2}. \par 
\section*{The PNG bias of the 21cm field}
The second ingredient required to study local primordial non-Gaussianity and its relation to the 21cm brightness temperature field is the parameter $b_\phi$, the other coefficient in Eqn.~\ref{eqn:tracerX}. We compute the PNG bias of the 21cm radiation field from its Separate Universe definition~\cite{Baldauf_2016, Barreira_2020,Barreira_2022}
\begin{equation}
    b_\phi^{21} = \frac{d \langle x_H (1+\delta_m) \rangle}{d\ln \sigma_8},
\end{equation}
where the derivatives are evaluated by running simulations with values of $\sigma_8 = \{0.79, 0.81, 0.83 \}$ analogously to Ref.~\cite{Barreira_2020}. The temperature normalization $T_0$ does not depend on $\sigma_8$, and since 21cm surveys measure the absolute brightness temperature and not its contrast we do not divide by the average signal in defining our bias parameters. \par 
We show our measurement of $b_\phi^{21}$, compared to $b_1^{21}$, in the bottom panel of Fig.~\ref{fig:b1}. Both measurements have been divided by $T_0(z)$ in order to be dimensionless. Our measured relation closely tracks prior results on the PNG bias of 21cm fluctuations from analytic excursion set calculations. Those calculations assume a modified barrier for the formation of an ionized region~(see Ref.~\cite{Lidz_2013} for details). They found $b_\phi$ to roughly follow the Universality relation of Ref.~\cite{Dalal_2008} with a modified collapse barrier
\begin{equation}
\label{eqn:bphi}
    b_\phi = \delta_B (b_1 - 1),
\end{equation}
where $\delta_B$ is close in value to the spherical collapse value $\delta_c = 1.686$. In Fig.~\ref{fig:b1} we plot Eqn.~\ref{eqn:bphi} assuming $\delta_B = 1.15$, and observe the relation matches the Separate Universe measurement until $z\approx 9$ and $\bar{x}_H \approx 0.2$, but begins to disagree at lower redshifts.
 \begin{figure}
    \centering
    \includegraphics[width=\columnwidth]{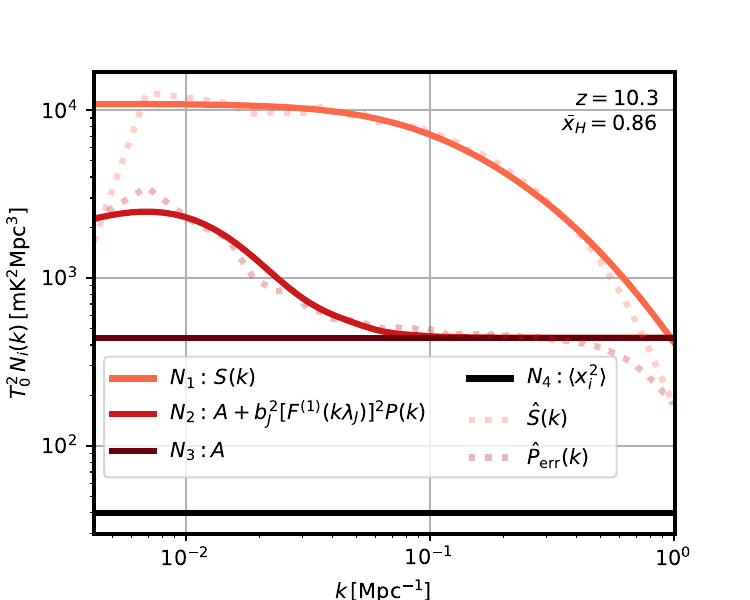}
    \caption{Different noise contributions to the 21cm power spectrum considered in this work at $z=10.3$, the snapshot where the zero-bias epoch occurs. The $N_i$ noises are defined in the Appendix. They span nearly three orders of magnitude in amplitude. Dotted lines correspond to the numerical estimates of noise to which smooth templates are fit.}
    \label{fig:noiseplots}
\end{figure}

 \section*{Effective noise in the 21cm power spectrum}
 \label{sec:noise}
Past studies of 21cm fluctuations at reionization have assumed that the only noise-like component which contributes to the covariance matrix of a forecast parameter constraint is the instrument's noise spectrum, which is a function of observed properties such as system temperature, observing time, collecting area and distribution of arrays (for interferometric observations). This is in contrast to galaxy surveys, which include a component related to the Poisson noise of the analysed galaxy sample. This noise is proportional to the number density of galaxies, $N \propto \bar{n}_g^{-1}$, and since the number density of hydrogen atoms is large their noise is neglected in 21cm forecasts. \par 
In the context of low-redshift HI intensity mapping surveys, it has recently been shown that neglecting this noise is not a good approximation~\cite{Villaescusa_Navarro_2018, Obuljen_2023, PhysRevD.110.063555}. Terms in the bias expansion of the 21cm intensity proportional to $\delta_m^2$ generate a white noise-like contribution, at large scales, to the power spectrum with amplitude 
\begin{equation}
\label{eqn:b2noise}
    N \sim \left. \langle \delta_m^2 (\bk) \delta_m^2 (\bk') \rangle'\right |_{k\to 0} \sim  \frac{b_2^2}{4} \int \frac{d^3 \bq}{(2\pi)^3} P_{mm}(q)^2,
\end{equation}
which is degenerate with sampling noise. For post-reionization HI intensity mapping this term was shown to be on the order of 10 times larger than the sampling noise present in HI-hosting halos. From Eqn.~\ref{eqn:bubbleexpansion} we expect this `$b_2$ noise' contribution to be present in the 21cm spectrum at reionization, being driven by the $b_2^x$ coefficient of ionization fraction fluctuations (and density-weighting of the linear bias $b_1^x$). However, this noise is not a fundamental floor, and techniques beyond a naive power spectrum analysis are expected to mitigate it~\cite{cabass2024cosmologicalinformationperturbativeforward}. \par  
We thus adopt four estimates of this `effective noise' noise from our simulation. These estimates of noise correspond to increasingly optimal survey analyses, culminating in noise level representative of field-level analysis which hit the sampling floor of 21cm radiation.  They are quantitatively defined in the Appendix. Our noise estimates are representative of:  
\begin{itemize}
    \item \textbf{Noise 1: }a naive analysis using the auto-power spectrum of 21cm fluctuations alone.
    \item \textbf{Noise 2: }an analysis which uses higher-order information (such as a bispectrum, or a field-level analysis), to mitigate the contribution of $b_2$ noise in the 21cm power spectrum. 
    \item \textbf{Noise 3: }similar to $\textbf{Noise 2}$ but where the influence of large-scale ionizing background fluctuations have also been mitigated. 
    \item \textbf{Noise 4: }an optimistic scenario where \emph{only} the sampling noise of the ionizing bubbles is present. 
\end{itemize}
We show these four noise estimates in Fig.~\ref{fig:noiseplots} for the zero-bias epoch in our simulation, $z=10.3$ and $\bar{x}_H = 0.84$.
        % deriv_fnl = (2*(b1*bphi*alpha + fnl*bphi**2 * alpha**2)*pk)

        % deriv_b1 = (2*(b1+fnl*bphi*alpha)*pk) 
\section*{The Fisher matrix}
For fiducial $f_{\rm NL}^{\rm loc} = 0$, we define the Fisher information on $f_{\rm NL}$ as~\cite{Castorina_2018} 
\begin{equation}
\label{eqn:fisher}
     F_{f_{\rm NL}}(z) = \frac{2 V(z)}{\pi^2} \int dk   \frac{k^2 \left ( b_1^{21} b_\phi^{21} \alpha(k)  \right )^2 P(k)^2}{\left ( (b_1^{21} )^2 P(k) + N_i(k) \right )^2},
\end{equation}
where $N_i(k)$, $i\in \{1,2,3,4\}$ is one of the four noise estimates described above. An experimental configuration sets $k_{\rm min}$ and $k_{\rm max}$, as well as an additional thermal noise $P_N (k)$. Since we are interested in assessing the ultimate viability of the zero-bias epoch we will neglect the presence of thermal noise.\par 
The Fisher matrix is calculated for several sub-surveys centered at the redshifts of our snapshots, spanning bandwidths of $\Delta \nu = 6 {\rm MHz}$. This bandwidth is a canonical bandwidth adopted by other 21cm studies at reionization~\cite{McQuinn_2006, Joudaki_2011} and strikes a balance between not encompassing too-wide of a redshift range (such that the properties of the brightness temperature field evolve significantly) and also being a frequency bandwidth over which foregrounds can be properly subtracted. Given a bandwidth, the largest scale probed is
\begin{equation}
    k_{\rm min} = 2\pi \frac{H(z)}{c (1 + z)^2} \frac{\nu_{\rm rest}}{\Delta \nu},
\end{equation}
assuming arbitrary large-scale transverse modes are measurable. The corresponding survey volume is given by 
\begin{equation}
    V(z) = 4\pi f_{\rm sky} \frac{\chi^2 (z) ( 1+z)^2}{H(z)} \frac{c \Delta \nu}{\nu_{\rm rest}}.
\end{equation}
\begin{figure}
    \centering
    \includegraphics[width=\columnwidth]{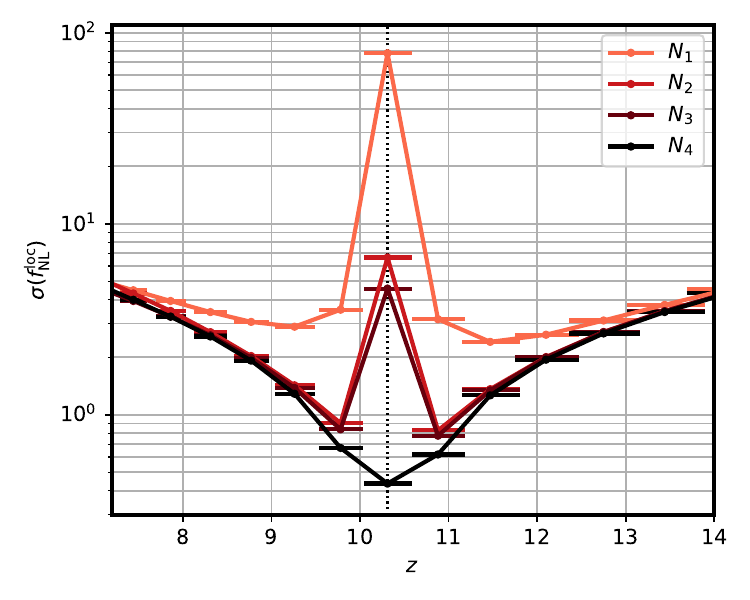}
    \caption{Forecasted uncertainty on  $f_{\rm NL}^{\rm loc}$ from 21cm surveys with bandwidths of $\Delta \nu = 6{\rm MHz}$, as a function of central redshift, near the zero-bias epoch in our fiducial \cmfast box. Dashed vertical line indicates the snapshot at which the bias of the 21cm fluctuation field crosses zero. The different lines labelled $N_i$ correspond to the four different characterizations of noise discussed in the text. Horizontal bars correspond to the redshift range covered by the $6$MHz bandwidth. }
    \label{fig:fisher}
\end{figure}
For $f_{\rm sky}=0.1$ this corresponds to survey volumes ranging between $V = 8-18 {\rm Gpc^3}$, and $k_{\rm min} \in \{0.07, 0.045\} {\rm Mpc}^{-1}$. The upper limit of this integral is set to $k_{\rm max} = 0.15 {\rm Mpc}^{-1}$, adopted in previous forecasts~\cite{Joudaki_2011}, as we are only interested in constraints on $f_{\rm NL}$ from large-scale modes where a bias expansion for the 21cm field is applicable~\cite{ McQuinn_2018}. For each snapshot we use bias parameters $\{b_1^{21}(z), b_\phi^{21}(z)\}$ extracted from our fiducial simulations, and shown in Fig.~\ref{fig:b1}. \par 
In Fig.~\ref{fig:fisher} we show, for these $\Delta \nu = 6{\rm MHz}$ surveys centered at various redshifts, the forecasted uncertainty $\sigma (f_{\rm NL}^{\rm loc}) = \sqrt{[F^{-1}]_{f_{\rm NL}}}$. We have marginalized over the linear bias $b_1^{21}$, to compare with past local PNG forecasts carried out with the 21cm power spectrum~\cite{Joudaki_2011, Lidz_2013}. Each line corresponds to a different noise spectrum, and assumes $f_{\rm NL} = 0$. \par 
Fig.~\ref{fig:fisher} shows the dramatic impact that $b_2$ noise has on our ability to constrain local PNG with a dense tracer like 21cm emission. Compared to past predictions, not accounting for $b_2$ noise can drastically degrade the available information at large scales. For snapshots neighbouring the zero-bias epoch, the degradation from $b_2$ noise can be as dramatic as a 5-fold increase in uncertainty. The constraints on local PNG are weakly sensitive to the central redshift of the 21cm survey if the naive $N_1$ analysis is carried out. The zero-bias epoch in this analysis is the \emph{worst} epoch to constrain PNG. Zero-bias tracers are only useful for surveys which are close to cosmic variance limited. When noise-limited, it is preferable to have a large linear bias (assuming $b_\phi$ is proportional to $b_1$)~\cite{Castorina_2018}. \par 
Turning to the analyses with $N_2$ and $N_3$, we see the improvement from mitigating $b_2$ noise in our ability to constrain PNG. Close to the zero-bias epoch, removing $b_2$ noise sharply increases the Fisher information. However, the residual $N_3$ noise in the error power spectrum is large enough that the zero-bias instant is not the maximally informative epoch. Nevertheless, redshifts near this epoch are the most constraining, and can cross the $\sigma (f_{\rm NL}^{\rm loc}) \sim 1$ threshold. We also notice, comparing $N_2$ and $N_3$ noise analyses, that the amount excess of scale-dependent noise present in our fiducial \opt simulation marginalyl degrades Fisher information.\par 
The $N_4$ analysis demonstrates what level of noise in the measurement is required to take advantage of said zero-bias epoch. The natural sampling floor of 21cm fluctuations at reionization is low enough that the Fisher information is highly enhanced when the linear bias of 21m radiation crosses zero. The uncertainty we find is on the order of $f_{\rm NL}^{\rm loc} \sim 0.4$ in this case, which is 50\% lower than the adjacent snapshots, and nearly a factor of 10 improvement over lower redshifts (which have historically been used for 21cm PNG forecasts). This simplified forecast highlights the significant improvement in constraining power from leveraging the zero-bias epoch of reionization to probe PNG. That this epoch occurs early in reionization ($\bar{x}_H \sim 0.2$) is encouraging, as the perturbative symmetries-based description we have adopted in this work breaks down late in reionization~\cite{McQuinn_2018}. \par 
Despite the simplicity of this forecast the redshift evolution of the different $N_i$ analyses in Fig.~\ref{fig:fisher} is non-trivial. In the Appendix we show that they can be understood analytically in certain limits. \par
\section*{Conclusions} In this work we have argued that, independent of the detailed astrophysics of reionization, there exists an epoch in which fluctuations in the brightness temperature of 21cm radiation are uncorrelated with the linear matter density field.\par 
We assessed whether the 21cm power spectrum at this epoch is a suitable probe of local PNG. We carried out a systematic analysis of the sources of noise-like contributions to the 21cm power spectrum which would reduce PNG constraining power, categorizing them in four different noise spectra $N_i$. The forecast constraints from these $N_i$ correspond to increasingly sophisticated analyses beyond the naive power spectrum, culminating in an analysis set by the shot noise in the ionization bubbles which dictate the 21cm signal. This characterization of noise was carried out using a simulation of reionization with the \cmfast code, with astrophysical parameters that produce a reionization scenario consistent with current datasets. We also note the zero-bias epoch has been indirectly observed in other simulations of reionization, at similar $\bar{x}_H$, despite completely different models for the astrophysics of reionization.\par 
We study how PNG forecasts vary for surveys situated at redshifts which span $z\in [7, 15]$, corresponding to when the Universe is 50\% to 10\% ionized by volume in our fiducial simulation, limited to scales where galactic foregrounds can be mitigated. We find that forecast uncertainties on local PNG at the zero-bias epoch, and the redshift range around it ($\Delta z \sim 1$), are significantly smaller (by factors of 10) than at other moments during reionization. However, we also find that improvements in constraining power around this epoch are strongly conditioned on effectively mitigating noise-like contributions arising from nonlinear biasing of the sources of reionization. Only analyses that truly hit the sampling floor of the 21cm power spectrum are capable of unlocking the full potential of PNG for this tracer. We stress that the methodology employed here is fully independent of simulation adopted, and should be used to further study the zero-bias epoch and its robustness as a probe of PNG in other astrophysical reionization scenarios.\par 
These results motivate a search for where this zero-bias epoch occurs in our actual Universe, once 21cm surveys begin to robustly measure this signal. This could be achieved, potentially, by cross-correlations between 21cm surveys and high-redshift galaxy surveys~\cite{La_Plante_2023, moriwaki2024insights21cmfield, fronenberg2024forecastsstatisticalinsightsline}. These results also motivate, in tandem, more sophisticated analyses than power spectrum-only for 21cm surveys in order to mitigate $b_2$ noise and reach the sampling floor of the 21cm radiation field. In the event of a detection of $f_{\rm NL}^{\rm loc}$ from current or future galaxy surveys~\cite{Sailer_2021, schlegel2022spectroscopicroadmapcosmic}, searching for the same signature in 21cm radiation will be an important cross-check and our results present a compelling experimental target for where to carry out such an analysis. \par
\section*{Acknowledgments} We thank Adam Lidz and Julian Muñoz for helpful discussions in the early stages of this work. We also thank Shi-Fan Chen and Jo Dunkley for enlightening conversations, and Gabriela Sato-Polito and Matias Zaldarriaga for additionally providing helpful comments on a draft of this work. NK acknowledges support from the Fund for Natural Sciences of the Institute for Advanced Study. Parts of this work were completed under the support of NSF award AST-2108126. Calculations and figures in this work have been made using \texttt{nbodykit} \citep{Hand_2018} and the SciPy Stack \citep{2020NumPy-Array,2020SciPy-NMeth,4160265}. This research has made use of NASA's Astrophysics Data System and the arXiv preprint server. The author is pleased to acknowledge that the work reported on in this paper was substantially performed using the Princeton Research Computing resources at Princeton University which is consortium of groups led by the Princeton Institute for Computational Science and Engineering (PICSciE) and Research Computing. 
\appendix

\section*{Appendix}
We present the four estimates of noise spectra adopted in this work. $N_1$ is defined as \emph{all} power present in the signal, which is uncorrelated with the linear density fluctuations. In the past literature it has been called the `stochasticity' $S(k)$: 
\begin{equation} 
\label{eqn:stoch}
\textbf{Noise 1: } S(k) = \langle \delta T_b | \delta T_b \rangle (k)  - \frac{\langle \delta T_b | \delta_m \rangle (k)^2}{\langle \delta_m | \delta_m \rangle (k)}.
\end{equation}
Measuring $S(k)$ in a simulation captures sources of non-Poisson noise such as `$b_2$ noise'. It was this definition of noise that Ref.~\cite{Castorina_2018} used to study their zero-bias tracers. In carrying out our forecasts, we fit to our measured $S(k)$ the perturbation theory template for `$b_2$ noise' given by the full $k$-dependent version of Eqn.~\ref{eqn:b2noise}, in order to suppress low-$k$ variance from our simulations. \par 
To disentangle $b_2$ noise from other components, we can also measure the ``error power spectrum" $\hat{P}_{\rm err}(k)$. It is defined as the spectrum 
\begin{equation}
    \hat{P}_{\rm err}(k) = \langle \hat{\epsilon}_{21} (\bk)\hat{\epsilon}_{21} (\bk')\rangle',
\end{equation}
where $\hat{\epsilon}_{21}(\bk)$ is a perturbative field-level estimate of the stochastic field in a bias expansion like Eqn.~\ref{eqn:tracerX}~\cite{Schmittfull_2019} 
\begin{equation}
\label{eqn:eps21heft}
    \hat{\epsilon}_{21} (\bk) = \delta T_b (k) - \sum_i \hat{b}_i \mathcal{O}_i (\bk). 
\end{equation}
In our case, $\mathcal{O}_i (\bk)$ are the advected bias fields defined in a Lagrangian bias expansion~\cite{Modi_2020, Schmidt_2021}. The best-fit biases $\hat{b}_i$ are obtained by minimizing the variance $\langle [\epsilon_{21}^2 ](\bx) \rangle$ as described at length in Refs.~\cite{Kokron_2022} and \cite{shiferaw2024uncertaintiesgalaxyformationphysics}. The shape of the error spectrum $P_{\rm err}$ is sensitive to neglected components in Eqn.~\ref{eqn:eps21heft}, such as higher-order perturbative contributions and fully non-perturbative effects. Increasing the number of bias operators $\mathcal{O}_i$ order-by-order should provide increasingly better estimates of the sampling error. Analyses which include statistics beyond the power spectrum (such as three-point functions, or full field-level analyses) will be fitting for these contributions and they will thus not be a form of noise.\par 
An explicit component which is missed by the standard symmetries-based bias expansion in $\delta_m$, used to generate Eqn.~\ref{eqn:eps21heft}, is a component which corresponds to the presence of large-scale fluctuations in ionizing (or H$_2$ dissociating) photon backgrounds~\cite{Wyithe_2009, Cabass:2018hum}. In Ref.~\cite{Cabass:2018hum} these effects were shown to organize themselves in a separate series expansion of terms, where the contributions to the biased density contrast can be described to lowest order as
\begin{equation}
\label{eqn:largebkgrnd}
    \delta_g \ni   F(k\lambda) (b_{J} \delta_m + \epsilon_J),
\end{equation}
where $\lambda$ is the mean free path of the radiation field, $b_{J}$ is an associated bias coefficient, $\epsilon_J$ is the stochasticity of the sources of this new background, and $F(k\lambda)$ is a scale-dependent function with functional form
\begin{equation}
    F(k\lambda) = \frac{{\rm arctan}(k \lambda)}{k \lambda}.
\end{equation}
For $k \lambda \gg 1$ we have $F(k \lambda) \propto (k\lambda)^{-1}$ and this term can be degenerate with the impact of $f_{\rm NL}^{\rm loc}$, and also adds power to large scales in a similar way to local non-Gaussianity. However, as this term is linear in $\delta_m$, the $N_1$ noise as defined by Eqn.~\ref{eqn:stoch} is insensitive to its presence. The presence of large-scale fluctuating radiation backgrounds in \cmfast, from Lyman-Werner photons (which suppress star-formation in molecular cooling galaxies) as well as ionizing, X-ray and Lyman-$\alpha$ photons that regulate star formation in cells~\cite{Qin_2020}, can induce fluctuations in the neutral hydrogen fraction with the form of Eqn.~\ref{eqn:largebkgrnd}. \par 
To account for these effects, at every snapshot, we fit a functional form 
\begin{equation}
\label{eqn:perr}
    P_{\rm err}(k) = A + F(k \lambda_J)^2 \left [  b_J^2 P(k) + B \right ],
\end{equation}
to the measured $\hat{P}_{\rm err}(k)$\footnote{This component cannot be easily included in a field-level -- $F(k\lambda)$ is nonlinear in $\lambda$, and the field-level estimator we employ relies on a linear dependence of the operator on the free coefficient.}. The constants $A$ and $B$ are an estimate of the shot noise of the ionization bubbles and ionizing sources, respectively, and the second term corresponds to this residual scale-dependent background term (with a bias $b_J$ and an effective mean free path $\lambda_J$). An estimate of the noise with both terms is denoted \textbf{Noise 2} $(N_2)$ and an estimate of the noise using only the inferred value of $A$ is \textbf{Noise 3} $(N_3)$. \par 
The pure sampling noise associated with $\delta T_b$ is difficult to quantify. It is very low in amplitude, and indeed is normally neglected in studies of 21cm fluctuations at reionization. Entertaining the possibility that our estimate of the sampling noise from Eqn.~\ref{eqn:perr} has not hit the sampling floor we will employ, as an ultimate test, an analytic estimate of the shot noise of 21cm fluctuations. In the bias expansion of ionization fraction fluctuations, Eqn.~\ref{eqn:bubbleexpansion}, the $\epsilon_x (\bx)$ term should be related to the abundance of ioniation bubbles. Given a mass function for bubbles $n(M)$ (such as what is calculated by Ref.~\cite{Furlanetto_2004}) the global ionization fraction is given by
\begin{equation}
    \bar{x}_i (z) = \int dM n(M) V(M),
\end{equation}
where $M$ is the mass enclosed in an ionized bubble of volume $V(M)$. This volume-weighted integral of the mass function also defines the effective bias of bubbles, where it weighs the different biases $b(M)$. The sampling noise for bubbles is calculated by assuming the ionization field is a random superposition of fully ionized spheres whose sizes are drawn from the mass function $dn/dM$. The $k\to0$ self-pair contribution, which defines the shot noise, is then given by the variance of their volumes~\footnote{This calculation is a conservative over-estimate. The assumption of spherical binary ionized bubbles introduces a profile $3 j_1 (kR_i)/kR_i$ to each bubble's Fourier transform which we've taken to be 1 in the $k\to 0 $ limit. }
\begin{align}
\textbf{Noise 4: }    P_{\rm shot, x} &= \frac{\langle (\epsilon^x)^2 \rangle}{\bar{x}_i^2}  \\
    \label{eqn:xhshot} &= \frac{1}{\bar{x}_i^2} \int dM n(M) V^2 (M),
\end{align}
in analogy with calculations from halo models for, e.g., intensity mapping~\cite{Schaan_2021}. We calculate the integrals in Eqn.~\ref{eqn:xhshot} using the analytic mass function of Ref.~\cite{Furlanetto_2004}, but re-scale their contribution using the global ionization fraction evolution from the \opt simulations we are analyzing. Calculating Eqn.~\ref{eqn:xhshot} as a function of redshift provides the $N_4$ noise curve used in the main text. \par 
In summary, $N_1$ includes flat noise and `$b_2$ noise', $N_2$ does \emph{not} include `$b_2$ noise' but is sensitive to scale-dependent $F(k)$ noise, $N_3$ is an estimate of the flat contribution to the $N_2$ curve, and $N_4$ is an analytic estimate from the theory of the distribution of ionized bubble sizes for the flat spectrum, to be contrasted with $N_3$. \par 
\begin{figure}
    \centering
    \includegraphics[width=\columnwidth]{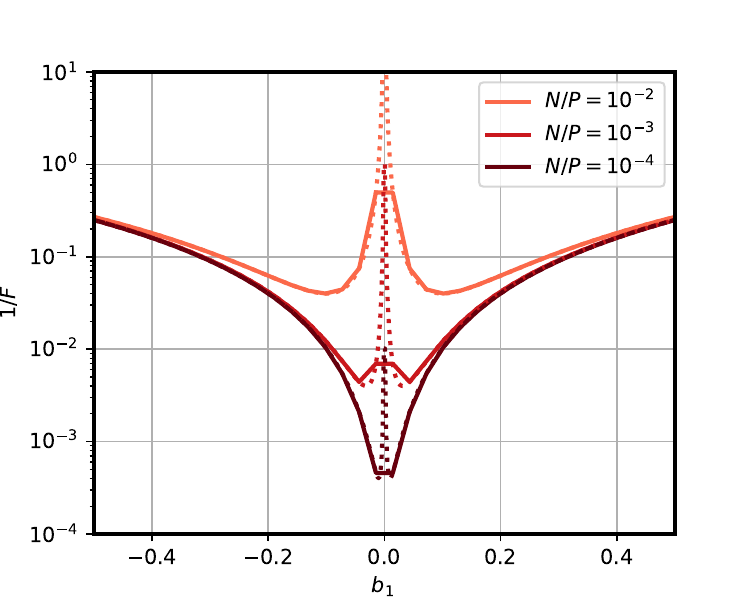}
    \caption{Simplified Fisher information assuming all time evolution is derived from $b_1$ only. The solid line shows a sparse sampling of points in the interval $b_1 \in \{ -1, 1\}$ of 70 points, comparable to the number of simulated snapshots in that interval, and the dotted lines show the equivalent expression under much denser sampling. The Fisher information evolves rapidly between the zero-bias moment and the Fisher maxima. The qualitative structure of the solid curves is strikingly similar to Fig.~\ref{fig:fisher}. }
    \label{fig:simplefisher}
\end{figure}
\section*{Analytic structure of the Fisher information} The Fisher matrix expressed in Eqn.~\ref{eqn:fisher} and the corresponding redshift-dependent structure of the forecast $\sigma (f_{\rm NL}^{\rm loc})$ in Fig.~\ref{fig:fisher} can be understood analytically from some simplifying arguments. For one, the monotonic evolution of $b_1^{21}(z)$ means we can also plot $F_{f_{\rm NL}}$ as a function of only $b_1^{21}$, and so the value of the 21cm linear bias sets the `clock' for the forecast. For a single wavenumber $k$, neglecting marginalization over $b_1$, the Fisher information is written as a Laurent polynomial in $b_1$
\begin{equation}
\label{eqn:simplefisher}
    F_{f_{\rm NL}} = \frac{A}{b_1^2} \frac{1}{\left ( 1 + \frac{B}{b_1^2} \right )^2}, 
\end{equation}
where $A$ is an amplitude $A = V(z(b_1^{21}))\, \Delta k k^2 \alpha(k)^2 b_\phi^2$ and $B = \frac{N}{P}$ is a dimensionless parameter that relates the noise amplitude at a $k$ scale to the amplitude of the \emph{matter} power spectrum at that $k$. This Fisher information has extrema at three points, $b_1^* = \{ -\sqrt{B}, 0, \sqrt{B} \}$. The points at $b_1^{21} = \pm \sqrt{B}$ are maxima, but the point at $b_1^{21} = 0$ is a \emph{minimum}. This is not surprising -- the original zero-bias argument in Castorina et al~\cite{Castorina_2018} assumes that $b_1^2 P \gg N$ at the same time that $b_1 \to 0$. For finite $N$ these conditions cannot be simultaneously satisfited, but it is true that the Fisher information diverges in $N\to 0$ (which can also be seen from all of the extrema $b_1^*$ going to the same value). $A$ is some amplitude that sets the overall scaling, but it is the value of $B$ for an analysis setup that sets what is the \emph{optimal} epoch for an $f_{\rm NL}^{\rm loc}$ search, near the zero-bias one, as well as how much additional benefit does carrying out a survey there compare to another. In Fig.~\ref{fig:simplefisher} we plot the reciprocal of Eqn.~\ref{eqn:simplefisher} for $A=1$ and various values of $B$, for two sampling schemes in $b_1$ ($N_b = \{70, 1000\}$ linearly spaced values in the interval $b_1 \in \{-1, 1\}$). Despite neglecting the full range of $k$-values, and any redshift evolution in $A(z)$ or $B(z)$ we see that Fig.~\ref{fig:simplefisher} possesses very similar structure to Fig.~\ref{fig:fisher}. The effect of finite sampling in $b_1$, from the limited number of snapshots is to over-estimate the exact zero-bias Fisher information. However, since surveys must always average over some redshift range, we take our fiducial survey setups centered at a simulation snapshot (with constant properties within that snapshot) to be a reasonable approximation. We also note in passing that the observed asymmetry in the heights / peak locations of Fig.~\ref{fig:fisher} come from the dependence of $b_\phi (b_1)$, as well as the evolution of the volume $V(z)$. Assuming $b_\phi \propto (b_1 - 1)$ in Eqn.~\ref{eqn:simplefisher} shifts the location of the maxima to $\{ -\sqrt{B}, \sqrt{B} \} \to \{-B-\sqrt{B^2 + B}, \sqrt{B^2 + B} - B\}$, and the value of $F$ is no longer the same at these maxima. \par

\bibliography{apssamp}
\end{document}